\documentclass[aps,pra,twocolumn,superscriptaddress]{revtex4-2}

\usepackage{graphicx}     
\usepackage{amsmath}      
\usepackage{amssymb}      
   
\usepackage{color}        
\usepackage{physics}      
\usepackage{siunitx}     
\usepackage{bm}
\usepackage{hyperref}

\newcounter{algoline}
\newenvironment{algorithm}[1][htbp]
  {\begin{table}[#1]
   \centering
   \refstepcounter{table}
   \renewcommand{\thetable}{Algorithm \arabic{table}}
   \begin{minipage}{0.9\linewidth}
   \small
   \setcounter{algoline}{0}
  }
  {\end{minipage}\end{table}}

\newcommand{\algstep}[1]{%
  \stepcounter{algoline}%
  \noindent\textbf{Step \thealgoline:} #1\par
}
\newcommand{\algfor}[2]{%
  \stepcounter{algoline}%
  \noindent\textbf{Step \thealgoline:} For #1 to #2 do\par
  \hspace*{1em}%
}
\newcommand{\algend}{\noindent\hspace*{1em}\textbf{end}\par}

\begin{document}

\preprint{APS/123-QED}

\title{
    Efficient Preparation of Fermionic Superfluids in an Optical Dipole Trap through Reinforcement Learning }

\author{Yueyang Min}
\affiliation{State Key Laboratory of Surface Physics, Key Laboratory of Micro and Nano Photonic Structures (MOE), and Department of Physics, Fudan University, Shanghai 200433, China}
\affiliation{Shanghai Qi Zhi Institute, AI Tower, Xuhui District, Shanghai 200232, China} 
\thanks{Y.M. and Z.L. contributed equally to this work.}

\author{Ziliang Li}
\affiliation{State Key Laboratory of Surface Physics, Key Laboratory of Micro and Nano Photonic Structures (MOE), and Department of Physics, Fudan University, Shanghai 200433, China}
\affiliation{Shanghai Qi Zhi Institute, AI Tower, Xuhui District, Shanghai 200232, China}
\thanks{Y.M. and Z.L. contributed equally to this work.}

\author{Yi Zhong}
\affiliation{State Key Laboratory of Surface Physics, Key Laboratory of Micro and Nano Photonic Structures (MOE), and Department of Physics, Fudan University, Shanghai 200433, China}

\author{Jia-An Xuan}
\affiliation{State Key Laboratory of Surface Physics, Key Laboratory of Micro and Nano Photonic Structures (MOE), and Department of Physics, Fudan University, Shanghai 200433, China}

\author{Jian Lin}
\thanks{jlin17@fudan.edu.cn} 
\affiliation{School of Physical Science and Technology, Ningbo University, Ningbo, 315211, China}
\affiliation{State Key Laboratory of Surface Physics, Key Laboratory of Micro and Nano Photonic Structures (MOE), and Department of Physics, Fudan University, Shanghai 200433, China}

\author{Lei Feng}
\thanks{leifeng@fudan.edu.cn}
\affiliation{State Key Laboratory of Surface Physics, Key Laboratory of Micro and Nano Photonic Structures (MOE), and Department of Physics, Fudan University, Shanghai 200433, China}
\affiliation{Institute for Nanoelectronic devices and Quantum computing, Fudan University, Shanghai, 200438, China}
\affiliation{Shanghai Key Laboratory of Metasurfaces for Light Manipulation, Shanghai, 200433, China}
\affiliation{Hefei National Laboratory, Hefei 230088, China}

\author{Xiaopeng Li}
\email{xiaopeng\underline{ }li@fudan.edu.cn}
\affiliation{State Key Laboratory of Surface Physics, Key Laboratory of Micro and Nano Photonic Structures (MOE), and Department of Physics, Fudan University, Shanghai 200433, China}
\affiliation{Shanghai Qi Zhi Institute, AI Tower, Xuhui District, Shanghai 200232, China}
\affiliation{Institute of Nanoelectronics and Quantum Computing, Fudan University, Shanghai 200433, China}
\affiliation{Shanghai Artificial Intelligence Laboratory, Shanghai 200232, China}
\affiliation{Shanghai Research Center for Quantum Sciences, Shanghai 201315, China}

\date{\today}

\begin{abstract}

We demonstrate a reinforcement learning (RL)–based control framework for optimizing evaporative cooling in the preparation of strongly interacting degenerate Fermi gases of $^6$Li. Using a Soft Actor-Critic (SAC) algorithm, the system autonomously explores a high-dimensional parameter space to learn optimal cooling trajectories. Compared to conventional exponential ramps, our method achieves up to 130\% improvement in atomic density within a 0.5 second, revealing non-trivial control strategies that balance fast evaporation and thermalization.
While our current optimization focuses on the evaporation stage, future integration of other cooling stages, such as grey molasses cooling, could further extend RL to the full preparation pipeline. Our result highlights the promise of RL as a general tool for closed-loop quantum control and automated calibration in complex atomic physics experiments.

\end{abstract}
\maketitle
\medskip 
\paragraph*{Introduction.---} 
Substantial advances have been made in the development and control of quantum systems, which have yielded numerous practical applications in recent decades. With the increasing complexity of quantum platforms, including atomic and molecular optics~\cite{Wigley2016amo,Tranter2018amo,Pilati2019amo,Davletov2020amo,Wu2020amo,Casert2021amo,Vendeiro2022amo,Anton2024amo,Xu2024amo,Reinschmidt2024amo}, quantum information~\cite{Ding2019qi,An2019qi,Baum2021qi,Lin2022qi,Mukherjee2025qi}, nanophotonics~\cite{Nadell2019np,Zhu2025np}, ultrafast laser engineering~\cite{Lu2020ul,Kokhanovskiy2019ul},experimental imaging and signal processing~\cite{Guo2021ip,Picard2020ip,Ness2020ip}, 
their peak performance depends on finely tuned optimization across high-dimensional, nonlinear, and strongly coupled parameter spaces.
One great example is the production of Bose-Einstein condensates (BEC)~\cite{Ketterle1999bec,Dalfovo1999bec}or degenerate Fermi gases~\cite{Granade2002dfg,Gross2016dfg}. 
Such process involves
multiple cooling stages, including magneto-optical trap (MOT), compressed MOT (CMOT)~\cite{PETRICH1994cmot,DePue2000cmot}, gray molasses (GM) ~\cite{Boiron1996gm,Oien1997gm}, and evaporative cooling~\cite{DAVIS1995ec,Sackett1997ec}, and requires extensive manual tuning based on experimental experience and intuition using trial-and-error approaches.

To overcome such limitation and automate the optimization process, machine learning techniques have been increasingly employed.
Early efforts applied Bayesian optimization to reduce experimental trial counts in high-dimensional search spaces, enabling efficient tuning of evaporative cooling paths and rapid BEC generation~\cite{Wigley2016amo,Vendeiro2022amo,Xu2024amo}. Neural network--based models have also been used to navigate tens of control parameters in magnetic trap compression, revealing nonlinear dependencies that are difficult to capture analytically~\cite{Tranter2018amo}. These approaches demonstrate that data-driven methods can uncover efficient
but often unintuitive strategies, offering scalable solutions
towards optical control of complex quantum systems. 

Despite thier great improvement in autonomous optimization, these methods mostly are on static sampling or cost-function regression. Such approaches typically optimize a fixed set of control parameters by evaluating the performance at isolated points. Regarding temporally or sequentially dependent processes, such as evaporative cooling, these methods appear to be unnatural to the intrinsic dynamics.
In contrast, RL with its natural structure of state, action, and reward—is well-suited for dynamic, closed-loop control RL agents can continuously adapt their policies based on feedback \cite{Anton2024amo,Milson_2023}, enabling them to learn time-dependent control strategies robust across varying initial conditions and noise environments.

In this work, we implement the SAC~\cite{Lin2022qi} reinforcement learning framework to autonomously optimize the procedure to produce strongly interacting degenerate Fermi gas of $^6$Li atoms. The procedure spans multiple stages, including MOT, CMOT and evaporative cooling, and they are formulated to be a closed-loop and real-time optimization task. In this framework, the SAC agent receives feedback in the form of reward signals derived from absorption images reflecting the atomic density distribution and learns to generate high-performance control policies across a high-dimensional, continuous parameter space. 
Due to hardware limitations, this work focuses specifically on optimizing the evaporative cooling stage.
Our results shows the efficacy of SAC in solving complex continuous control tasks and future promises for broader applications.

\begin{figure}[htp]
\centering
\includegraphics[width=\linewidth]{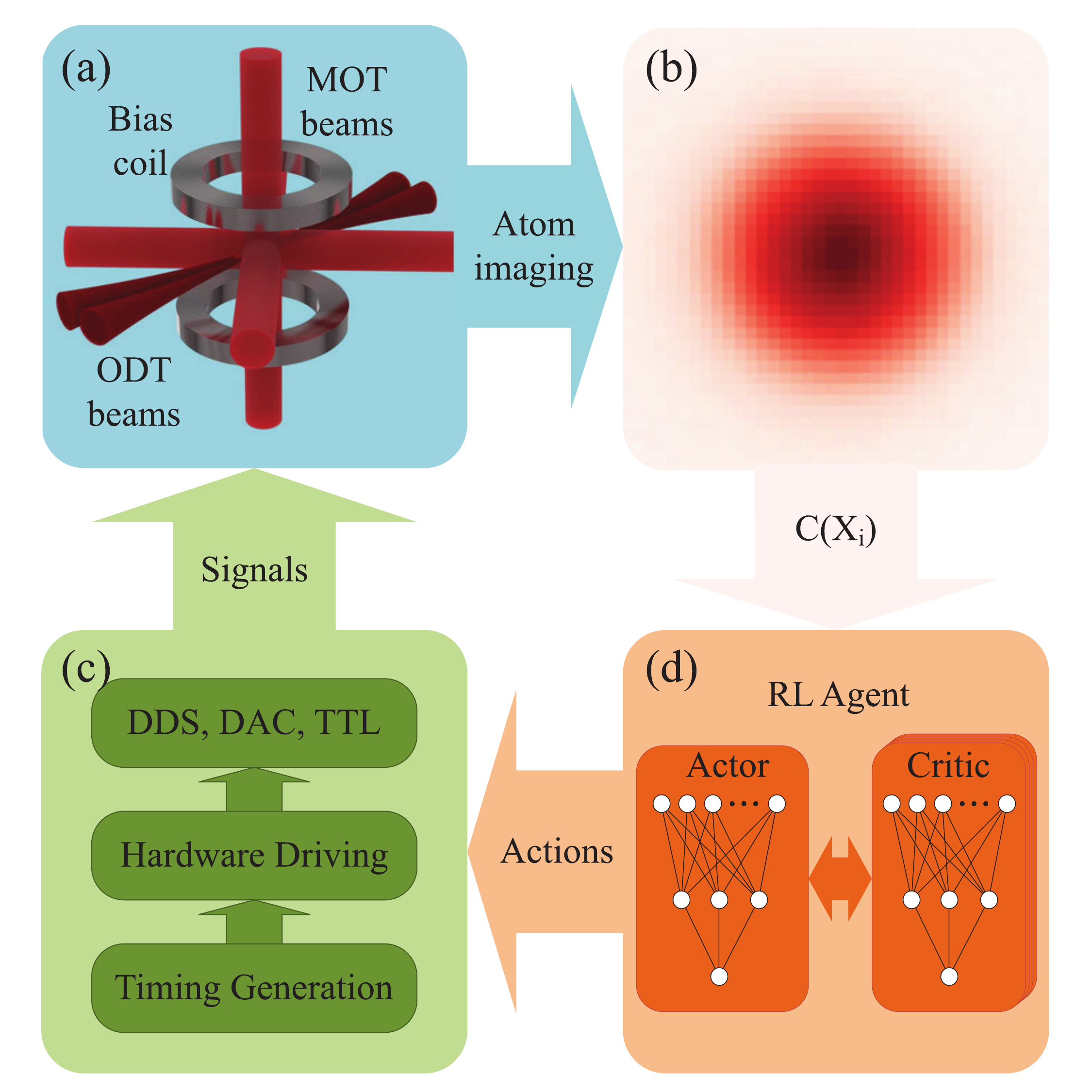}
\caption{
        Illustration of the reinforcement learning (RL) training procedure. 
        (a) Schematic diagram of the experimental setup, showing the MOT beams, two 1064~nm optical dipole trap (ODT) beams, and Helmholtz coils. 
        (b) Absorption image of the atomic cloud. 
        (c) Parameters proposed by the RL agent are encoded by the control system into physical control signals for hardware execution. 
        (d) Architecture of the Soft Actor-Critic (SAC) model used as the RL agent.
    }
\label{fig:setup}
\end{figure}

\medskip 
\paragraph*{Experimental Apparatus for Strongly Interacting Fermi Gas.---} 
The production of a degenerate Fermi gas of $^6$Li atoms consists of three main stages: MOT, CMOT, and evaporative cooling in a crossed optical dipole trap (ODT). The lithium oven is first heated to $365\,^\circ\mathrm{C}$. Hot atoms from the oven enter the Zeeman slower and decelerate to suitable velocities. A three-dimenssional MOT captures and futher cools the atoms. After loading and cooling in the MOT for 5~s, we transfer the atoms to a crossed optical dipole trap (ODT) by two $1064\,\mathrm{nm}$ laser beams with a beam waist of  $150\,\mu\mathrm{m}$ and crossing angle of $10^\circ$.
To improve transfer efficiency, a CMOT phase is applied immediately after the ODT is switched on. During this stage, the magnetic field gradient and laser detuning are dynamically adjusted to increase atom number in the ODT. At the end of the CMOT, the MOT quadrupole field is rapidly switched off. Meanwhile the repumping beam is turned off slightly earlier than the cooling beam to prepare the atoms in an approximate $1{:}1$ spin mixture of the hyperfine ground states $|1\rangle \equiv |F=1/2, m_F=+1/2\rangle$ and $|2\rangle \equiv |F=1/2, m_F=-1/2\rangle$.

Following successful loading into the ODT, a pair of Helmholtz coils are activated to ramp the magnetic field to $832\,\mathrm{G}$, close to the broad Feshbach resonance between $|1\rangle$ and $|2\rangle$. This enables strong interactions, with s-wave scattering length $1/a\approx0$,  
(unitary). 
and initiates the evaporative cooling process. By progressively reducing the ODT laser power, the atoms further cool down and form a degenerate Fermi gas in the end.

To optimize the critical evaporative cooling stage, we apply the reinforcement learning algorithm. In this framework, the evaporative cooling sequence is divided into three segments. Each segment is characterized by a set of control parameters, including time duration, power ramp profile, and the final power level, resulting in 20 parameters in total. Feedback or cost function to the algorithm comes from the observation of the atomic density distribution after time of flight (TOF).

In contrast to weakly interacting Bose gases, where
the condensate fraction can be easily extracted through a bimodal fitting of the TOF absorption imaging, 
it is nontrivial to extract the condensate fraction of strongly interacting Fermi gases from TOF imaging.
This presents a significant challenge in constructing an effective cost function for RL optimization. To overcome this difficulty, we adiabatically ramp the magnetic field to $690\,\mathrm{G}$ in about 50 ms. 
Through this adiabatic process to the BEC regime, the atomic Cooper pairs are converted to  Feshbach molecules. 
The ODT is then switched off, allowing the atoms to expand for $3\,\mathrm{ms}$, after which absorption imaging is performed on the $|2\rangle$ state. This measurement characterizes the entropy\cite{Stefano2008rmp} and gives the upper bound of the phase space density (PSD) of the original fermi gas.

PSD can serve as an effective cost function for our optimization because it is a fair trade-off between low temperature and high atom number
\cite{Anderson1995,Vendeiro2022amo}. In practice, we use a proxy of the PSD by evaluating the average atomic density within the cloud, 
\begin{equation}
n_{\text{avg}} = \frac{N_{\text{atom}}}{\sigma_x \sigma_y}, 
\label{eq:n_avg}
\end{equation}

\noindent with $\sigma_{x}$ ($\sigma_y$) the root-mean-square width of the atomic distribution, $n(x,y)$, along the $x$ ($y$) direction. 

\medskip 
\paragraph*{Implementation of Soft Actor-Critic Method.---}
The SAC reinforced learning consists of signal $\mathbf{s}_t$, action $\mathbf{a}_t$, and reward $\mathbf{r}_t$ at a specific time step $t$. The singal $\mathbf{s}_t$ is used to actuate the movement of the real system. In our experiment, it is a vector of the normalized 20 control parameters of the power ramp during evaporating cooling, with each element within $[-1, 1]$. The action $\mathbf{a}_t$ is randomly sampled according to policy $\pi$ at each time step $t$ based on the reward $\mathbf{r}_t$, namely $n_{avg}$ in Eq.~\ref{eq:n_avg}, via the reward function $\mathbb{E} _\pi[\sum_t(\gamma^tr_t)]$ with $\gamma\in(0,1]$ a discount factor. The feedback signal for the next move is then given by $\mathbf{s}_{t+1} = \mathbf{s}_t+\mathbf{a}_t$. Among various RL approaches for policy optimization, we focus on the entropy-regularized SAC method~\cite{haarnoja2018soft}, a state-of-the-art RL algorithm. Here the reward function is modified to 
$
\mathbb{E}_{\pi}\left[\sum_{t} \gamma^t \left( r(\mathbf{s}_t, \mathbf{a}_t) + \alpha \mathcal{H}(\pi(\cdot|\mathbf{s}_t)) \right)\right],
$
where \(\mathcal{H}\) represents the entropy of the policy \(\pi\), and \(\alpha\) is a relative weight that controls the trade-off between reward maximization and entropy, and is automatically tuned to stabilize the learning process. This formulation encourages the agent to learn stochastic policies that act as randomly as possible while achieving high task performance.

In practice, SAC employs an actor-critic architecture with deep neural networks, as shown in Fig.~\ref{fig:setup}(d). During SAC training, the input of each step is specified to be $\{t,\mathbf{s}_t\}$. The actor generates the action distribution under the guidance of the critic that employs two Q-networks and updates according to the temporal-different error via Polyak averaging ~\cite{polyak1992acceleration}with a target Q-network to mitigate overestimation bias. The agent chooses an action $\mathbf{a}_t$ by sampling from the actor network, and adds to $\mathbf{s}_t$. Here we set the action element $a_m\in[-0.05,0.05]$.

\begin{figure}[htbp]
\centering
\includegraphics[width=\linewidth]{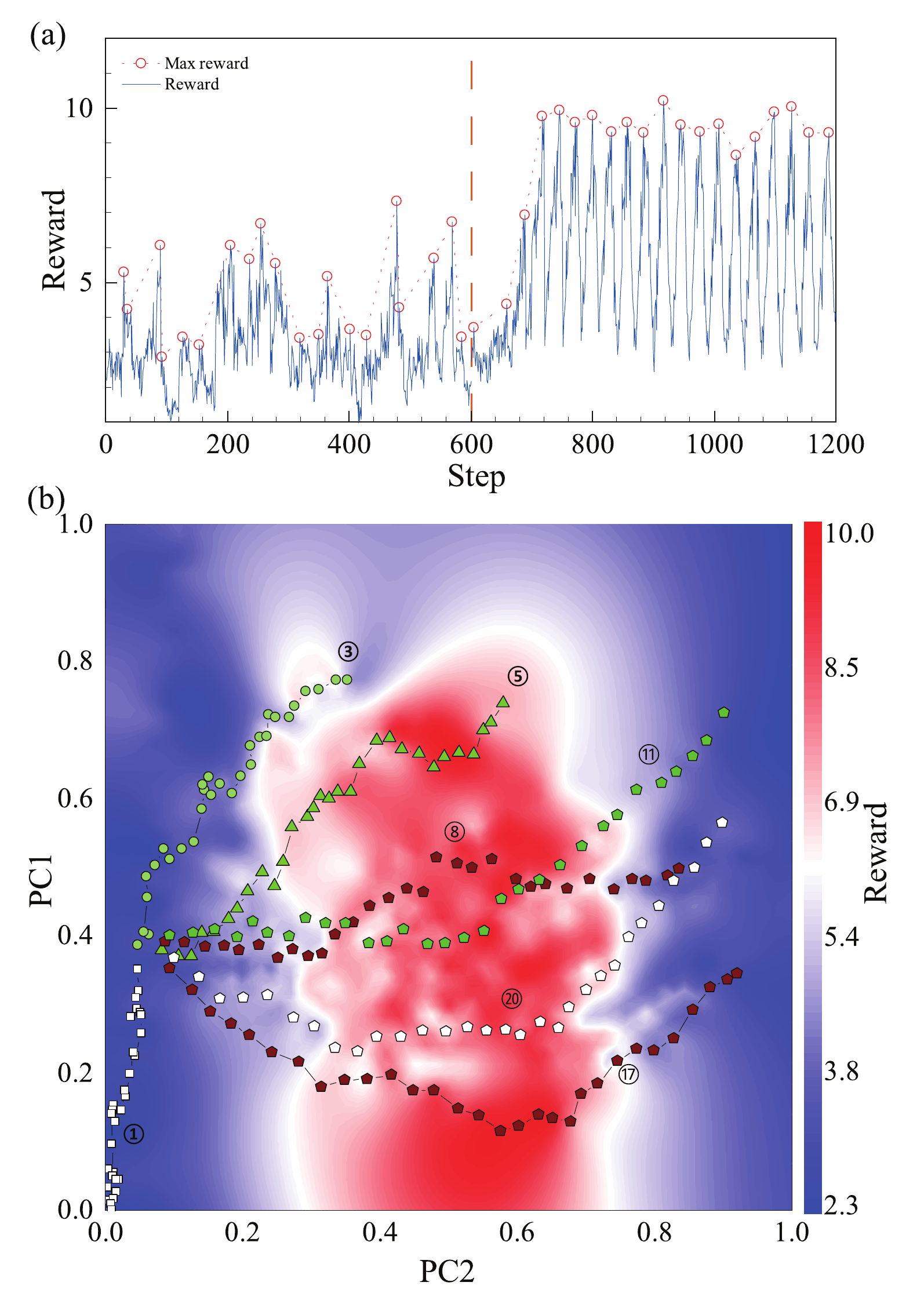}
\caption{
(a) Evolution of the cost function $n_{\text{avg}}$ across training episodes. The blue curve represents the SAC-guided optimization trajectory, while red circle indicate the best-performing configuration found in each episode. The vertical yellow line marks the boundary between the initial random exploration phase and the subsequent SAC-driven optimization phase.  
(b) Exploration trajectory in the control parameter space, projected via PCA for 600 SAC-guided trials (excluding initial random sampling). Each point corresponds to a tested parameter set, with color representing the associated $n_{\text{avg}}$ value. Background contours depict the interpolated reward landscape inferred from experimental data. Different marker shapes indicate distinct optimization episodes during SAC training.
}
\label{fig:evolution}
\end{figure}
\begin{figure*}[htp]
\centering
\includegraphics[width=\linewidth]{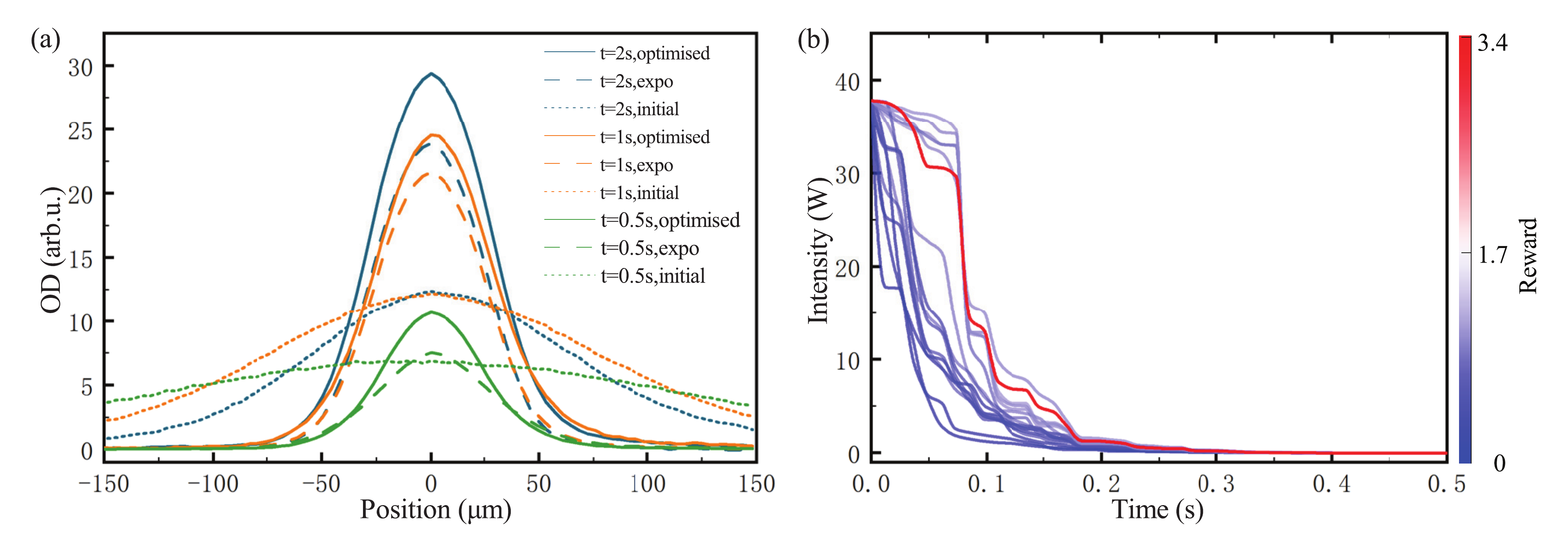}
\caption{(a) Integrated atomic density profiles (along one axis)
following optimized evaporation trajectories at different durations (0.5, 1, and 2\,s), compared with initial parameter settings and exponential ramps. 
(b) 
Time-dependence of the  optical power in the evaporative cooling process learned by SAC.
} 
\label{fig:results}
\end{figure*}

In our experiment, the RL optimization protocol comprises 40 episodes and each consists of 30 experimental iterative steps. During the first 20 episodes, the agent performs random exploration of the parameter space to populate the replay buffer with a diverse set of ramping trajectories. In subsequent episodes, the SAC algorithm is employed to refine the control policy based on accumulated experience.

Each episode begins from the same fixed initial configuration, which is a piecewise-linear optical power ramping trajectory. At each step, the RL agent proposes a normalized parameter update, which is then converted by the control system into physical signals, including direct digital synthesis (DDS) signals that drive the acousto-optic modulators (AOMs), digital-to-analog converter (DAC) and transistor-transistor logic (TTL) signals that control the magnetic field strength, optical trap intensity, and associated switching operations. After each episode, all system parameters are reset to their initial values to eliminate the influence of residual drifts and ensure consistent learning conditions.
The SAC agent performs asynchronous policy updates by sampling mini-batches from the replay buffer, thereby balancing exploitation of high-reward regions with ongoing exploration of the broader parameter landscape.

Following every experimental iteration, the signal-action-reward tuple $(\mathbf{s}_t, \mathbf{a}_t, \mathbf{r}_t)$ is recorded as a step in the replay buffer for subsequent SAC training.
For efficient exploration of the entire parameter space, the RL algorithm is calibrated so that any parameter point within the control domain can be reached within 30 iterative steps.

\paragraph*{SAC Performance on Cooling Strongly Interacting Fermions.---}
We first apply the SAC method to a slow evaporative cooling sequence of $2$~s, shown in Fig~\ref{fig:evolution}. In this figure, we summarize the optimization process that comprises $20$ randomly sampled episodes, namely $600$ steps followed by $20$ SAC-guided episodes. During the initial exploration phase (episodes $1$–$20$), the agent samples broadly across the parameter space, typically starting from suboptimal settings where the average density or reward $n_{\mathrm{avg}} \sim 2$ while occasionally discovering configurations yielding $n_{\mathrm{avg}} \approx 7$, shown in Fig.~\ref{fig:evolution}(a). After gathering approximately $600$ training samples, the SAC agent quickly identifies optimal ramp trajectories achieving $n_{\mathrm{avg}} > 8$ within just a few episodes and converges consistently to high-performance strategies. 
Because in each episode, the RL agent restarts at the same initial state, transient reward drops are observed at the beginning of each episode during the continuous training. The  algorithm setup is designed  to explore a broader parameter space and avoid convergence to local optima.
Nevertheless,
the agent reliably rediscovers near-optimal strategies before each episode ends.

To better visualize the optimization dynamics, 
we perform dimension reduction and project the 20-dimensional control parameter space onto a reduced 2-dimensional space through  principal component analysis (PCA). 
The projected PCA components are shown in Fig.~\ref{fig:evolution}(b).
Each marker corresponds to an optimization step within a particular episode, while the background contours depict the interpolated reward landscape derived from a large set of experimental data. Early exploration points (circles and squares) cluster in low-reward blue regions, whereas later SAC-guided steps gradually move toward high-reward red regions. The entropy-regulated policy encourages free exploration to avoid local optima, 
which accounts for the overshot in certain training episodes.

We further constrain the total evaporative cooling time ($t_{\rm evap}$) to $0.5$~s and $1$~s, and investigate whether RL could still successfully  produce a high-quality ultracold Fermi gas with a significant condensate fraction. 
This task is  more challenging than the $t_{\rm evap} = 2$~s case, as 
the system must traverse the phase transition from the high-temperature normal phase to the low-temperature superfluid within a shorter timescale.
At these timescales, the system already exhibits deviations from the quasi-static limit  \cite{Liu2021prl} . 
We find remarkable improvement in the condensate fraction in performing RL (Fig.~\ref{fig:results}). Especially for $t_{\rm evap} = 0.5$ s, although RL is initiated by a strategy producing no sizable condensate fraction, it develops an efficient convergence to  an optimized evaporative cooling strategy that  produces a fermionic superfluid with a significant condensate fraction. 
In comparison to the empirical exponential ramp widely used in ultracold atom experiments~\cite{Anderson1995,Makuto2003pra}, we find an enhancement of 130\% (19\%) in the final condensate fraction, 
for $t_{\rm evap} = 0.5$~s ($2$~s).
Despite that faster ramps in evaporative cooling causes higher atom lass rate due to lack of sufficient equilibration, RL converges to  a  meticulously design of non-adiabatic control for improving $n_{\rm avg}$.  
The significant improvement at shorter evaporative cooling times highlights the strength of RL in approaching optimal control of the strongly interacting Fermi gas.

In addition to enabling low-temperature superfluid production within a limited time window, the RL-designed evaporative cooling protocol exhibits conceptual novelty beyond the conventional exponential ramp. 
We show the optimized optical power profile for the case of $t_{\rm evap} = 0.5$~s in Fig.~\ref{fig:results}. The SAC-optimized curve reveals three distinct dynamical stages: 
(i) an initial \textit{slow descent}, slower than exponential;  
(ii) a \textit{rapid nonlinear drop} interspersed with a series of plateaus, which we expect to facilitate local thermalization and energy redistribution in the atomic sample;  and 
(iii) a final \textit{saturation stage} where the last 2\% of power is removed over  more than  one  half of the total time budget to accommodate the critical slowing down near the superfluid phase transition.

\medskip 
\paragraph*{Discussion.---}
Our results demonstrate that RL–assisted evaporative cooling significantly outperforms traditional exponential strategies in preparing unitary Fermi gases, especially under non-equilibrium conditions. The learned nonlinear trajectories achieve up to a 130\% improvement in the cost function $n_{\text{avg}}$ for a 0.5-second evaporation sequence compared to exponential ramps, highlighting the autonomous capability of RL in discovering  nontrivial and practical strategies without prior knowledge. This performance advantage is primarily attributed to two key features of the SAC framework: (i) efficient exploration of the high-dimensional control parameter space without the need of explicit models, and (ii) entropy-regularized policy that balances exploration toward unknown parameter space and exploitation of past experience and reward, which is especially beneficial for identifying metastable plateaus near quantum phase transitions.

Due to the lack of direct thermometry in the unitary regime, we characterize temperature of the degenerate Fermi gas by analyzing the condensate fraction in the BEC region. The values of the measured temperature over fermi temperature, $T_\text{mole} /T_F$ , are 0.26, 0.27, and 0.38 for evaporation sequences with durations of 2 s, 1 s, and 0.5 s, respectively. The measured temperature $T_\text{mole}$ represents the upper bound of the actual temperature in the unitary regime becuase the entropy-conserving transfer to the BEC side elevates the measured temperature\cite{Haussmann2007pra}.

The significant improvements of evaporative cooling within very short duration by RL sheds light towards fast repetition of  ultracold atom experiments. 
The current limitations of the optimization process are primarily imposed by the present  hardware constraints in our experiments. One significant constraint comes from limited bandwidth of the magnetic field adjustments. 
Given faster and more accurate field control, we can incorporate the MOT and CMOT stages into our SAC method to further improve the atomic density while reducing cooling time. Besides, the integration of grey molasses cooling and improved ODT geometries could further enhance evaporative cooling efficiency. 
We expect fast production  of degenerate Fermi gas exceeding a $1$~Hz repetition rate is within the optimization scope of SAC method, once the entire cooling process is let under control by SAC.

\paragraph*{Acknowledgement.---}
This work is supported by the Innovation Program for Quantum Science and Technology of China (Grant No. 2024ZD0300100),
the National Basic Research Program of China (Grants No. 2021YFA1400900), 
Shanghai Municipal Science and Technology (Grant No. 25TQ003, 2019SHZDZX01, 24DP2600100), National Natural Science Foundation of China (Grant No.12304555). 

\nocite{*}

\bibliography{reference}

\renewcommand{\thefigure}{S\arabic{figure}}  
\renewcommand{\thetable}{S\arabic{table}}   
\renewcommand{\thesection}{S\arabic{section}}
\setcounter{figure}{0}  
\setcounter{table}{0}
\setcounter{section}{0}

\clearpage  
\title{Supplementary Information for Efficient Preparation of Fermionic Superfluids in an Optical Dipole Trap through Reinforcement Learning}
\maketitle

\section{Experimental Sequence}
\label{app:sequence}

\begin{figure*}[hbt]
\centering
\includegraphics[width = \textwidth]{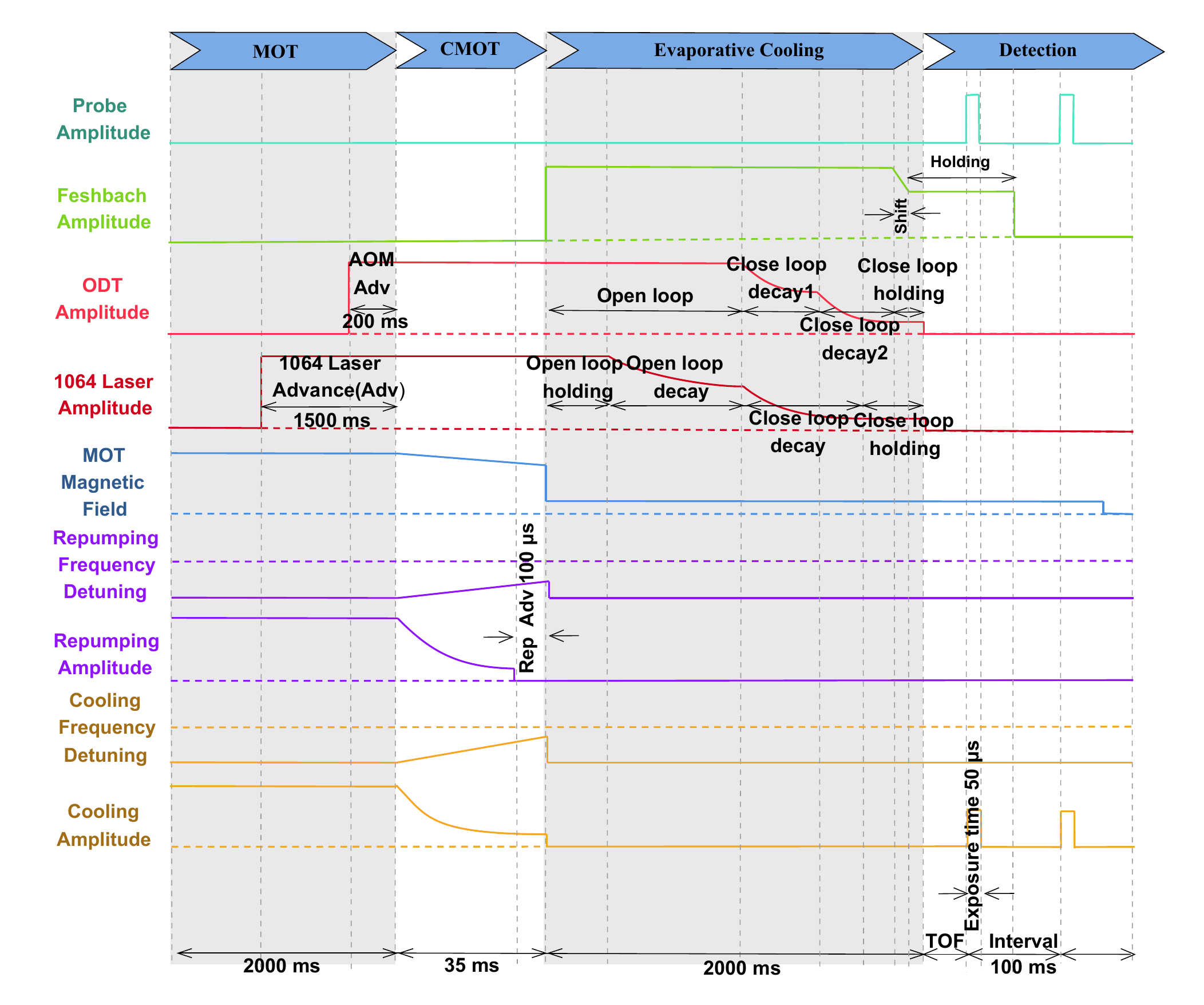}
\caption{
Experimental timing sequence. Full timeline of major control signals: DDS frequencies for cooling/repump lasers (orange/purple), DAC amplitudes for magnetic field (blue / green) and optical trap power (red), and TTL pulses for imaging (cyan). All timings are synchronized to $\pm\SI{10}{ns}$ via FPGA.
}
\label{fig:sequence}
\end{figure*}

The experimental sequence for preparing degenerate Fermi gases, as referenced in the main text, comprises four stages: MOT, CMOT, evaporative cooling, and detection (Fig.~\ref{fig:sequence}). During the MOT stage, the cooling beams operate at a detuning of $-10\Gamma$ ($\Gamma = 2\pi \times \SI{5.87}{MHz}$ being the natural linewidth) with an intensity of $17I_0$ ($I_0 = \SI{2.54}{mW/cm^2}$ denoting the saturation intensity), while the repump light is detuned by $-6\Gamma$. Transitioning to the CMOT stage, the cooling beam detuning is linearly increased to $-4\Gamma$ with a reduced intensity of $0.034I_0$, and the repump detuning adjusted to $-2\Gamma$, under magnetic field gradients of $\SI{20}{G/cm}$ (vertical) and $\SI{10}{G/cm}$ (horizontal). The evaporative cooling stage implements the optimized power curve, decreasing the dipole trap power from $\SI{37}{W}$ to $\SI{4}{mW}$. Final detection employs resonant probe light at $0.03I_0$ with a $\SI{50}{\mu s}$ exposure time. This procedure cools up to $8 \times 10^4$ atoms to quantum degeneracy after a $\SI{2}{s}$ evaporative cooling starting from a loaded mot of $6 \times 10^8$ atoms.

\section{Optimization Parameters}
The evaporative cooling trajectory is parametrized into three distinct stages, each governed by hardware-specific saturation limits. For the $i$-th stage, the optical intensity $I_i(t)$ follows:
\begin{equation}
I_i(t) = I_i^\text{start} + (I_i^\text{end} - I_i^\text{start}) \cdot f(t/t_i; \mathbf{p}^{(i)})
\end{equation}
where $I_i^\text{start}$ and $I_i^\text{end}$ denote the stage-specific boundary intensities, $t_i$ is the stage duration, and $f(\tau;\mathbf{p}^{(i)})$ is a normalized shape function ($f(0)=0$, $f(1)=1$) uniquely determined by an $m$-dimensional parameter vector $\mathbf{p}_i$. This parametrization decouples hardware constraints from learnable dynamics, enabling efficient RL optimization while respecting physical limits. 

The shape function $f(\tau;\mathbf{p}^{(i)})$ is constructed via a sampling-interpolation scheme that enforces both smoothness and monotonicity. For an $m$-parameter implementation, we uniformly sample $m$ interior points at $\tau_j = j/(m+1)$ ($j=1,\ldots,m$), assigning $f(\tau_j)=a_j$ with boundary conditions $f(0)=0$, $f(1)=1$. The values $a_j$ are recursively defined to ensure monotonic trends:
\begin{equation}
a_j = a_{j-1} + (1 - a_{j-1}) p_j, \quad p_j \in [0,1], \quad a_0 \equiv 0,
\end{equation}
where $\mathbf{p}^{(i)} = (p_1,\ldots,p_m)$ constitutes the learnable parameter vector. A cubic convolution interpolant then generates the full function $f(\tau)$, guaranteeing $\mathcal{C}^2$ continuity while preserving the monotonic property. This parametrization spans diverse curve families (exponential, polynomial, etc.) through its $[0,1]$-normalized parameters. 

\begin{algorithm}
\caption{Parametric Intensity Trajectory Generation}
\label{alg:intensity}

\algstep{Input: Stage index $i$, intensities $I_i^{\mathrm{start}}$, $I_i^{\mathrm{end}}$}
\algstep{Input: Duration $t_i$, parameters $\bm{p}_i \in [0,1]^m$}
\algstep{Output: Profile $I_i(t)$ for $t \in [0,t_i]$}

\algstep{Initialize $a_0 \gets 0$ \quad (Anchor at $\tau=0$)}
\algfor{$j = 1$}{$m$}
  \algstep{Compute $a_j \gets a_{j-1} + (1 - a_{j-1}) \cdot p_j$}
\algend
\algstep{Set boundary $a_{m+1} \gets 1$ \quad (Anchor at $\tau=1$)}
\algstep{Sample points $\{(\tau_j,a_j)\}$ where $\tau_j = j/(m+1)$}
\algstep{Interpolate $f(\tau) \gets \mathrm{CubicConvInterp}(\{(\tau_j,a_j)\})$}
\algstep{Compute $I_i(t) \gets I_i^{\mathrm{start}} + (I_i^{\mathrm{end}} - I_i^{\mathrm{start}}) \cdot f(t/t_i)$}
\end{algorithm}

The complete optimization space comprises 20 parameters: 15 for the stage curves (3 stages $\times$ 5 parameters), 3 stage durations, plus the dipole trap loading time and post-evaporation hold time. 

\begin{figure}[htp]
\centering
\includegraphics[width=\linewidth]{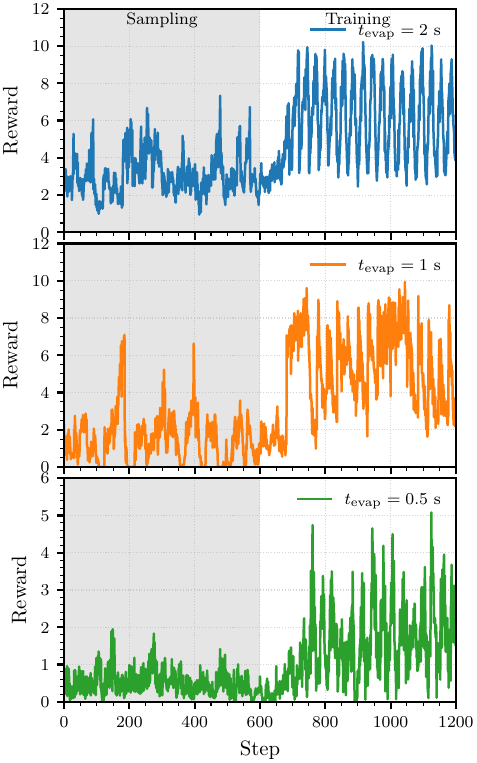}
\caption{
Full optimization trajectories for evaporation durations of $\SI{2}{s}$ (top), $\SI{1}{s}$ (middle), and $\SI{0.5}{s}$ (bottom). Gray area represent 600 random samples during initial exploration, while others show 600 SAC-guided optimization steps. The vertical axis displays the reward $n_\text{avg}$ (phase-space density proxy).
}
\label{fig:progress}
\end{figure}

\section{Optimization Dynamics}  
The SAC algorithm demonstrates remarkable optimization capability, efficiently navigating high-dimensional parameter spaces. As shown in (Fig.~\ref{fig:progress}), for all evaporation durations ($t_\text{ODT} = 0.5, 1,  \SI{2}{s}$), the agent begins exploration in low-reward regions, with 600 random samples populating the replay buffer. Notably, the $t_\text{ODT} = \SI{0.5}{s}$ case exhibits the strongest initial bias, as all of early samples cluster are near the starting point with rewards $n_\text{avg} < 2$.  Despite this handicap, SAC identifies near-optimal configurations within 5-7 episodes.

\begin{figure}[htpb]
\centering
\includegraphics[width=\linewidth]{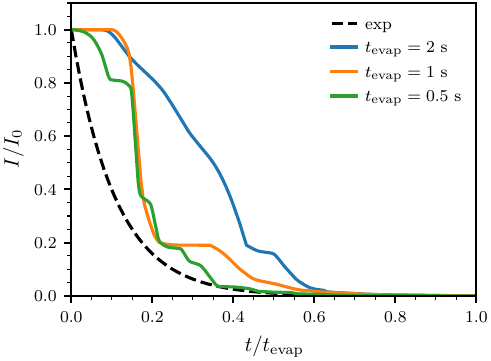}
\caption{
RL-optimised evaporate curve (normalized). $t_\text{evap}=\SI{2}{s}$ (blue), $\SI{1}{s}$ (orange), and $\SI{0.5}{s}$ (green). The black dashed line shows the exponential ramp curves as a benchmark baseline.
}
\label{fig:evapcurve}
\end{figure}

The optimized evaporation trajectories exhibit consistent nonlinear profiles across all durations (Fig.~\ref{fig:evapcurve}), systematically deviating from exponential ramps through two key features: First, all RL-designed curves demonstrate an initial gradual descent during the early evaporation stage, facilitating efficient atom loading and thermalization. Second, they transition into accelerated power reduction in the intermediate regime before finally approaching a near-adiabatic slope near degeneracy. This pattern becomes increasingly pronounced at shorter durations—the $\SI{0.5}{s}$ trajectory spends over $\SI{60}{\%}$ of its total time in the final slow-descent stage, compared to just $\SI{40}{\%}$ for the $\SI{2}{s}$ case, indicating sensitivity to non-adiabatic losses. The universal preference for this shaped trajectory suggests an optimization principle that balances rapid thermal atom removal with preservation of quantum degeneracy.

\section{Verification of Quantum Degeneracy}
\label{sec:BEC_verification}

To verify the quantum degeneracy of the strongly interacting Fermi gas after evaporative cooling, we ramp the magnetic field to the BEC regime. Here the atomic cloud is characterized through absorption imaging at $\SI{690}{G}$, near the BEC limit of the BEC-BCS crossover. The emergence of a bimodal distribution in $\rho_\text{2D}$ unambiguously confirms molecular BEC formation, demonstrating cooling below the degeneracy threshold. To extract the optical depth (OD) distribution, we acquire three consecutive images separated by $\SI{100}{ms}$: (i) atomic shadow $I_\text{atom}$, (ii) probe intensity $I_\text{probe}$, and (iii) background $I_\text{background}$. The OD is calculated as:
\begin{equation}
\text{OD}(x,y) = \ln \left( \frac{I_\text{probe}(x,y) - I_\text{background}(x,y)}{I_\text{atom}(x,y) - I_\text{background}(x,y)} \right),
\end{equation}
with the column density $\rho_\text{2D}(x,y)$ obtained via:
\begin{equation}
\rho_\text{2D}(x,y) = \frac{\text{OD}(x,y)}{\sigma_0}\left( 1+\frac{4\Delta^2}{\Gamma^2} + \frac{I_\text{probe}}{I_\text{sat}} \right),
\end{equation}
\noindent where $\sigma_0$ is the resonant cross-section $\sigma_0=3\lambda^2/2\pi $, and $\Delta$ accounts for imaging detuning. 

\begin{figure*}[b]
\centering
\includegraphics[width=\textwidth]{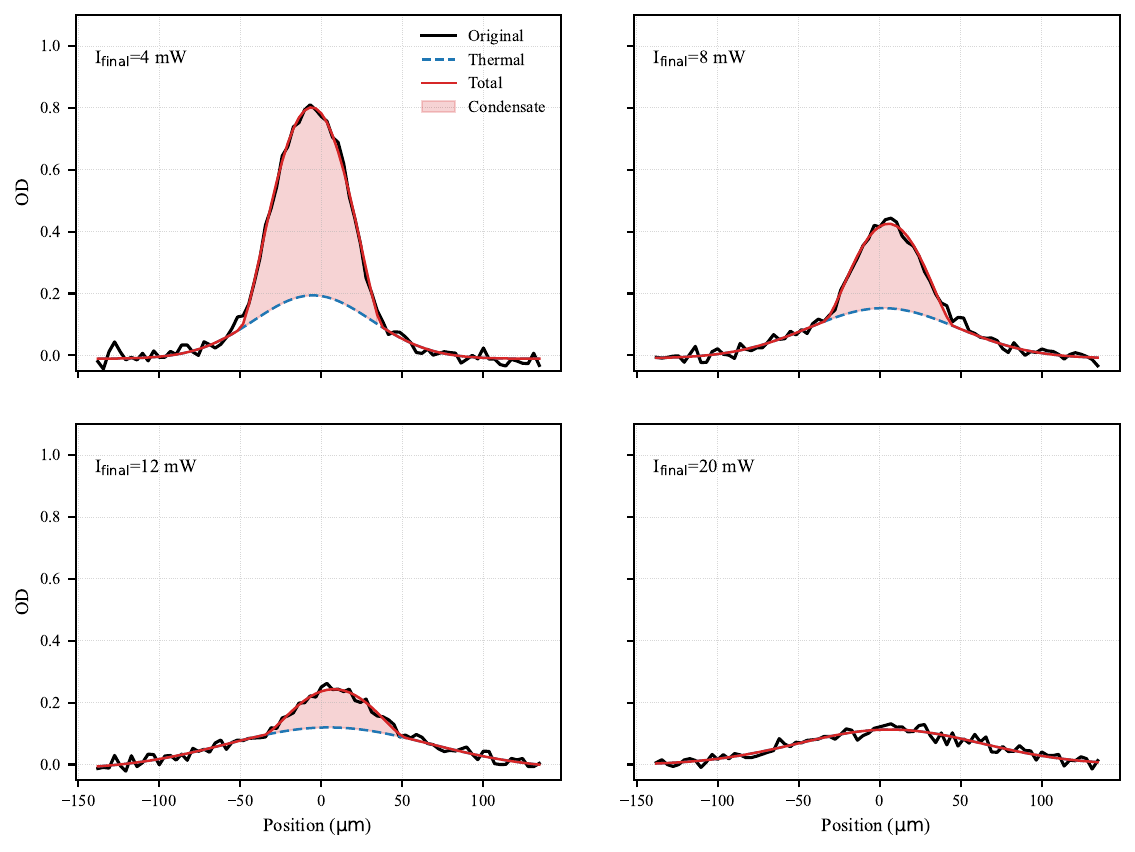}
\caption{
Evolution of molecular BEC formation for varying final optical dipole trap powers. Shown are the axial optical depth profiles (black solid curves) with bimodal fits (red solid) decomposing into thermal (blue dashed) and condensed (light red shaded) components. At $\SI{20}{mW}$, the atomic cloud exhibits purely thermal distribution. Progressive condensation emerges below $\SI{12}{mW}$, culminating in $\SI{4}{mW}$ with $\SI{76}{\%}$ condensate fraction.
}
\label{fig:bimode}
\end{figure*}

To quantitatively characterize the evaporative cooling process, we stop at various times during the evaporation to measure the formation of the molecular BEC. We choose to observe at the times when the ODT intensity $I_\text{final} = 20, 12, 8$ and $\SI{4}{mW}$ (Fig.~\ref{fig:bimode}). 
At $\SI{20}{mW}$, the atomic cloud exhibits a purely thermal Boltzmann distribution. Progressive reduction to $\SI{12}{mW}$ initiates BEC onset, with the density profile developing a central parabolic peak superimposed on the thermal Gaussian background. This bimodal structure intensifies at lower trap depth, culminating in our optimized condition where the condensate fraction reaches $76\%$. This threshold behavior validates the SAC-optimized trajectory's choice of $I_\text{final} = \SI{4}{mW}$.

The atomic cloud temperature was determined through condensate fraction. We separate the thermal and condensed components by fitting the column density distribution $\rho_{\text{2D}}(x,y)$ using a double-Gaussian function,  The condensate fraction $N_0/N_\text{mole}$ was then extracted from the integrated atom numbers. The condensate fraction relates to reduced temperature via:
\begin{equation}
\frac{N_0}{N_{\text{mole}}} = 1 - \left( \frac{T_{\text{mole}}}{T_c} \right)^3,
\end{equation}
where $T_c$ is the critical temperature given by:
\begin{equation}
T_c = \frac{\hbar \bar{\omega}}{k_B} \left( \frac{N_{\text{mole}}}{\zeta(3)} \right)^{1/3} \approx 0.42 \, T_F.
\end{equation}
\noindent 
Here $\bar{\omega} = (\omega_x \omega_y \omega_z)^{1/3}$ is the geometric mean trap frequency, $N_{\text{mole}}\approx N_\text{atom}/2$ the molecule number, $\zeta(3) \approx 1.202$ the Riemann zeta function, and $T_F$ the Fermi temperature. We measure $T_{\text{mole}}/T_F\approx0.26$ for $\SI{2}{s}$ evaporate trajectory, $T_{\text{mole}}/T_F\approx0.27$ and $0.28$ for $\SI{1}{s}$ and $\SI{0.5}{s}$ trajectories respectively using the same method.

\end{document}